# Compressibility and thermal expansion of $YBa_2Cu_3O_x$

K. Grube[*], H. Leibrock[*], W. H. Fietz[*], S. I. Schlachter[*], A. I. Rykov[†], S. Tajima[†], P. Schweiss[‡], B. Obst[*], and H. Wühl[§*]

*Forschungszentrum Karlsruhe, [*]ITP and [‡]INFP, 76021 Karlsruhe, Germany*
[†]*Superconductivity Research Laboratory, ISTEC, Tokio 135, Japan*
[§]*Universität Karlsruhe, IEKP, 76128 Karlsruhe, Germany*

*The linear compressibilities and the thermal expansion of an untwinned, nearly optimal doped $YBa_2Cu_3O_{6.94}$ single crystal were measured along the three crystallographic axes with a high-resolution dilatometer mounted in a high-pressure cell. The measurements were performed in a temperature range from 50 K to 320 K under hydrostatic gas pressure of maximal 0.65 GPa. At a temperature of 300 K the measured bulk modulus is reduced and shows a very strong enhancement under increasing pressure due to pressure-induced oxygen ordering. This ordering process can also be related to a glass transition seen in the thermal expansion above 225 K. At lower temperatures, when oxygen ordering is frozen, the bulk modulus is increased and shows a significantly reduced enhancement under pressure. This pressure dependence, however, is still anomalous high, probably caused by bonds of extremely different compressibilities within the unit cell. The widely spread literature data of the bulk modulus can be explained by this high pressure dependence.*
    PACS numbers: 62.20.Dc, 64.70.Pf, 74.72.Bk

The compressibility of high-$T_{\rm C}$ superconductors has been studied extensively since their discovery with different experimental methods. However, in contrast to simpler materials like copper or NaCl, there is no correspondence between the derived results. For $YBa_2Cu_3O_x$ the literature data of the reciprocal of the compressibility, the bulk modulus, at ambient pressure and room temperature scatters from 40 GPa to 180 GPa.[1,2] The values seem to depend on the maximal achieved pressure of each experiment: So, most of the ultrasonically determined data at ambient pressure are far below the values measured in gas-pressure cells at maximal 2 GPa and these are again smaller than the results of measurements performed in diamond-anvil cells under pressures up to 200 GPa. Provided these discrepancies are not a result of experimental problems they point to an anomalous high pressure dependence of the bulk modulus $dB/dP$ which is indeed reported by several authors.[1] Since most of the measurements on $YBa_2Cu_3O_x$ were performed at room temperature a possible explanation for an intrinsic origin of a high $dB/dP$ value is the thermally assisted rearrangement of oxygen atoms in the CuO chains[3] which can occur at oxygen concentrations nearly up to $x = 7.0$ and temperatures above 240 K. This oxygen ordering process can also be induced by pressure[4,5] because the unit-cell volume shrinks with increasing order.[6,7] To prove whether this additional volume reduction leads to a smaller bulk modulus with an anomalous high pressure dependence we investigated the length changes of a high-quality $YBa_2Cu_3O_x$ single crystal with a high-resolution dilatometer at temperatures between 50 K and 320 K under hydrostatic pressure up to 0.65 GPa.

The crystal was grown by a pulling technique,[8] detwinned and cut into a cube of about $3 \times 3 \times 3$ mm. After its oxygenation an oxygen content of $x = 6.94$ was determined by a neutron-diffraction study, also, revealing a degree of twinning of less than 1 %. In a subsequent microchemical study by energy dispersive x-ray spectroscopy (EDX) no signs of impurities were found within the detection limit of less than 0.1 % .

The measurements were carried out by a capacitive dilatometer which we have miniaturized and mounted in a gas-pressure cell. The thermal expansion and bulk modulus of the sample are measured by recording the sample-length change under variation of temperature and helium-gas pressure, respectively. The resolution of the dilatometer $\Delta L/L < 10^{-8}$ is high enough to determine the bulk modulus directly without using any equation of state. The relative temperature control is better than 1 mK neglecting the unavoidable temperature variation of maximal 0.2 K due to pressure application. Pressure can be adjusted within $\pm 0.01$ MPa enabling us to separate compressibility and thermal-expansion effects from other length changes.

We studied the thermal expansion of the crystal between 50 K and 320 K at ambient pressure and 0.49 GPa. At the transition temperature $T_c = 92$ K all linear thermal-expansion coefficients exhibit clear discontinuities. Under pressure no shift of $T_c$ was detectable corresponding to a slightly overdoped $YBa_2Cu_3O_x$ crystal.[9] Beyond 225 K we found the signature of a glass transition (see Fig. 1a) which we relate to the evolving oxygen-ordering process. This glass transition shows a positive pressure dependence indicating that the time constant $\tau$ of the ordering process grows with increasing pres-sure. This corresponds to measurements from Tissen et al.[5] who studied a YBa$_2$Cu$_3$O$_{6.41}$ crystal in a larger pressure range and reported an exponential increase of $\tau$.

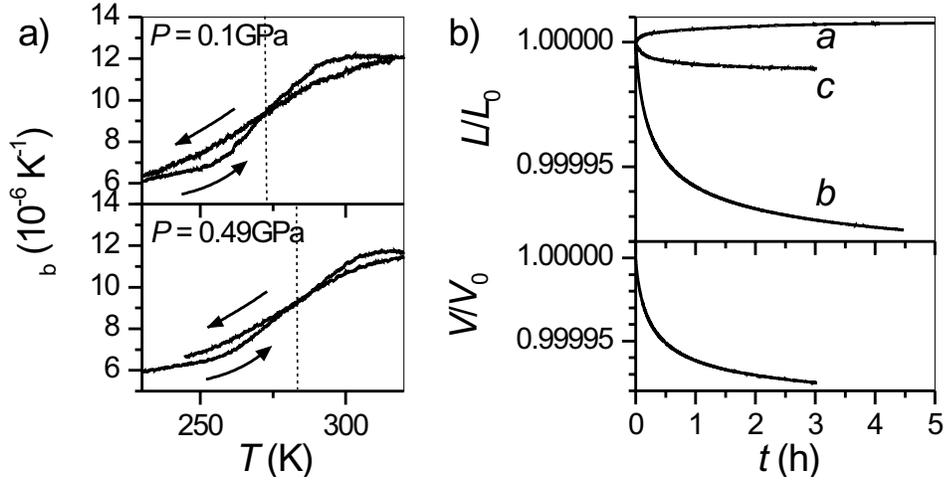

Fig. 1. a) The glass transition seen in the linear thermal-expansion coefficient along the b axis of YBa$_2$Cu$_3$O$_{6.94}$ at ambient pressure and 0.49 GPa measured during cooling and heating with a rate of ±8 mK/s. b) The relative lehngth of the a, b, and c axis and the relative volume of YBa$_2$Cu$_3$O$_{6.94}$ versus time after the pressure of 0.65 GPa was applied at a constant temperature of 300 K.

In order to detect the shrinkage of the unit-cell volume caused by the ordering process we first heated the sample to a temperature of 300 K, waited until thermal equilibrium was reached, and applied then a pressure of 0.65 GPa. Keeping temperature and pressure constant we recorded the length change during the next hours. Figure 1b shows the measured length and volume changes as a function of time after pressure was applied. The a axis relaxes to an slightly enhanced value whereas the b and c axis, as well as the entire volume, reveal qualitatively an exponential reduction with increasing time. The largest decrease is revealed by the b axis although transition-temperature measurements prove that in this direction the length of the CuO chains is enlarged by the oxygen-ordering process because additional charge carriers are produced.[3–5]

In order to study the influence of oxygen ordering upon the bulk modulus and its pressure derivative we measured the linear compressibilities of the lattice parameter during pressure release beginning with 0.55 GPa at a constant temperature of 300 K. We chose the slowest possible pressure-release rate of $-0.02$ MPa/s to maintain as far as possible thermal equilibrium conditions. However, since even this slow rate could not provide thermal equilibrium in the entire pressure range, the resulting bulk modulus values give an upper limit, only. For comparison with literature values we performed a linear fit to our bulk modulus data, corresponding to a Murnaghan equation of state, and get at ambient pressure $B_0 = 99.7$ GPa with a tremendous pressure dependence of $dB/dP = 20$. According to the relaxation of the lattice parameter at constant pressure the pressure dependence of the linear compressibility along the $b$ axis has the strongest influence upon $dB/dP$, too. The high $dB/dP$ value originates in the volume reduction caused by the oxygen-order process because the achieved degree of order depends on temperature, pressure and the laboratory time scale. As the time constant of the ordering process increases exponentially under pressure[5] the crystal leaves thermal equilibrium if the pressure is increased high enough. This results in an apparent hardening of the structure because the rearrangement of oxygen atoms is hindered kinetically.

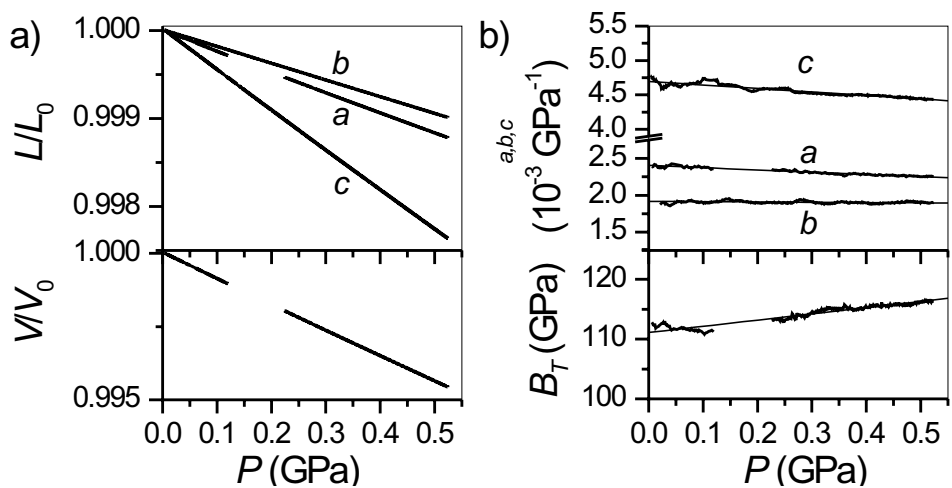

Fig. 2. a) The original data of the relative length of $YBa_2Cu_3O_{6.94}$ along the $a$, $b$, and $c$ axis and the relative volume as a function of pressure at a constant temperature of 200 K. b) The resulting linear compressibilities, the bulk modulus, and the linear fits to them versus pressure.

To prove if this high pressure dependence of the bulk modulus is caused by oxygen order only, we repeated the measurements at a temperature of 200 K where the ordering process is frozen. Consequently, we detected no relaxation effects. The data of the compressibility measurements at 200 K are presented in Fig. 2a and b. A linear fit to the compressibility data

## Compressibility of YBa$_2$Cu$_3$O$_x$

results at ambient pressure in a linear compressibility of the $a$, $b$, and $c$ axis of $2.40 \cdot 10^{-3}$ GPa$^{-1}$, $1.92 \cdot 10^{-3}$ GPa$^{-1}$, and $4.70 \cdot 10^{-3}$ GPa$^{-1}$ with a pressure dependence of $-3.0 \cdot 10^{-4}$ GPa$^{-2}$, $-0.4 \cdot 10^{-3}$ GPa$^{-2}$, and $-5.1 \cdot 10^{-4}$ GPa$^{-2}$, respectively (see Fig. 2b). For the bulk modulus we obtained $B_0 = 111.2$ GPa and a pressure dependence of $dB/dP = 10$. In comparison with the result at 300 K, $B_0$ is significantly higher which can not be assigned to normal hardening due to the lower temperature but is caused by the lack of oxygen ordering. As expected, at 200 K, $dB/dP$ is considerably reduced. In comparison to usual materials, however, it is still extremely high, indeed, pointing to an additional intrinsic feature which leads to a rapid rise of the bulk modulus with increasing pressure. At 200 K the main part of $dB/dP$ is based on the high pressure dependence of the linear compressibility along the $c$ axis direction. We suppose that the very soft bonds between the copper ions of the CuO$_2$ planes and their adjacent apical oxygen ions in $c$ direction produce this anomalous high $dB/dP$ because under small pressures the major part of the resulting volume change arises from the compression of these bonds. (For a detailed explanation of this effect see Ref. 11). Furthermore, upon contraction of these bonds under pressure, they transfer charge from the CuO chains to the CuO$_2$ planes.[11] Due to their ionic character a change in their charge balance can alter considerable the electrostatic attraction leading to a reduction of the bond lengths and, thus, a further hardening.

In conclusion, our measurements show that the widely spread literature data of the compressibility of YBa$_2$Cu$_3$O$_x$ can be explained by an enhanced pressure dependence of the bulk modulus. This is caused by a very high pressure dependence of the linear compressibility along the $c$ axis, probably, because of the soft bonds between the copper atoms of the CuO$_2$ planes and the apical oxygen atoms. At temperatures above 240 K the pressure dependence of the bulk modulus is further increased by an oxygen-ordering process in the CuO chains which is hindered kinetically with increasing pressure.

This work was partially supported by the New Energy and Industrial Technology Development Organization in Japan.